# Suspension flow: do particles act as mixers?

A. Boschan*[a], M.A.Aguirre [a], G. Gauthier [b]

a Grupo de Medios Porosos, Facultad de Ingeniería, Universidad de Buenos Aires, Argentina.
E-mail: abosch@fi.uba.ar

b Laboratoire Fluides Automatique et Systèmes Thermiques, Campus universitaire d'Orsay, France.

**Recently, Roht et al. [J. Contam. Hydrol. 145, 10-16 (2013)] observed that the presence of suspended non-Brownian macroscopic particles decreased the dispersivity of a passive solute, for a pressure-driven flow in a narrow parallel-plates channel at low Reynolds number. This result contradicts the idea that the streamline distortion caused by the random diffusive motion of the particles increases the dispersion and mixing of the solute. Therefore, to estimate the influence of this motion on the dispersivity of the solute, and investigate the origin of the reported decrease, we experimentally studied the probability density functions (pdf) of the particle velocities, and spatio-temporal correlations, in the same experimental configuration. We observed that, as the mean suspension velocity exceeds a critical value, the pdf of the streamwise velocities of the particles markedly changes from a symmetric distribution to an asymmetric one strongly skewed to high velocities and with a peak of most probable velocity close to the maximum velocity. The latter observations and the analysis of suspension microstructure indicate that the observed decrease in the dispersivity of the solute is due to particle migration to the mid-plane of the channel, and consequent flattening of the velocity profile. Moreover, we estimated the contribution of particle diffusive motion to the solute dispersivity to be three orders of magnitude smaller than the reported decrease, and thus negligible. Solute dispersion is then much more affected by how particles modify the flow velocity profile across the channel, than by their diffusive random motion.**

# Introduction

Particle suspension flows are present in numerous industrial and natural situations, at different length scales[1], and are of considerable importance in many chemical, hydrocarbon, and environmental processes. Also, the mixing and dispersion of dissolved species, such as a passive solute, in subsurface confined flows, is of relevance in waste storage and water management applications. The two fields have rarely been brought together: simultaneous transport of particles and solutes has been studied by a number of authors [2-3], mostly with a focus on the differential transport, and breakthrough curves of the two. It is known that whether due to size exclusion effects [4], or to shear enhancement mechanisms [5], more frequently colloids are less dispersed than solutes in simple confining geometries. However, less attention has received the effect of non-Brownian suspended particles on the dispersion of solutes [6-7].

Regarding the latter, recently Roht et al.[7] have observed a decrease of the dispersivity of the solute in the presence of particles, for flow between parallel-plates at low Reynolds number. It is frequently considered that the diffusive random motion[6,8] exhibited by the particles, due to the long-range hydrodynamic interactions among them, is a source of dispersion enhancement for the solute, with the particles acting as mixers. The result by Roht et al. [7] contradicts this idea. Moreover, it has been shown in numerical simulations [9-10] that, in a parallelepiped channel, a suspension may organise itself in layers and the particles close to the channel wall remain almost at rest[11]. The increased rugosity of the channel resulting from these particles might also enhance solute mixing[12].

Nevertheless, at low particle Reynolds number, particles might undergo shear-induced diffusion [13-14] due to many-body collisions or contact (in the case of rough particles) [15-16]. As a consequence, under inhomogeneous shear, suspension organises itself: particles migrate from high to low shear regions and this migration leads to a flattening of the suspension velocity profile as reported by Lyon and Leal [17] for a pressure driven flow between parallel plates. This behavior was also predicted by theoretical and numerical studies[18-20]. In simple flow configurations, this flattening would more likely hinder than enhance global dispersion mechanisms of the solute, e.g. Taylor dispersion, which arises from the combination of strong velocity gradients and molecular diffusion in the wall-normal direction. This occurs, for example, in the flow of shear-thinning polymer solutions between parallel-plates or in capillary tubes. In these situations, the velocity profile flattens if compared with that of the Newtonian solvent, and it has been shown theoretically and experimentally [21-22] that this generates a reduction in the dispersivity of a dissolved solute.

In short, through some mechanisms, particles are expected to enhance mixing and dispersion of the solute, while through others, they are expected to reduce it. To determine the relative influence of the competing effects, and shed light on the origin of the decrease reported by Roht et al. [7], we studied the suspension microstructure and particle velocity statistics in the same experimental configuration.

## Experimental setup

Suspensions were made of monodisperse polystyrene spherical particles of density $\rho = 1.05$ g/cm$^3$ and radius $a = 20 \pm 1$ μm. Particles were suspended in a Newtonian water glycerol mixture (~21% in weight), to achieve a neutrally buoyant suspension of particle volume fraction $\phi = 0.05$. At a room temperature of 24 °C, no buoyant displacement was observed after 4 days of inspection for any of the suspensions used in the experiments. In our estimation, this implies that a possible buoyant displacement of the particles, during our longest experiments, would be smaller than one sixth of the channel thickness. To avoid aggregation effects, a small amount of SDS surfactant was added.

The channel (Fig. 1), horizontally set, is constructed with two rectangular parallel flat glass plates separated by two mylar strips that also sealed the longest sides of the glasses. Its dimensions were 250 mm long (streamwise), 80 mm wide (spanwise) and $420 \pm 10$ μm thick (wall-normal). A reservoir supplies suspension to the channel inlet; axial flow to the outlet is established by means of a syringe pump sucking the suspension out of the channel at constant flow rate. To avoid transient effects, data is acquired only after a lapse of time equivalent to that required for a particle to transverse the channel.

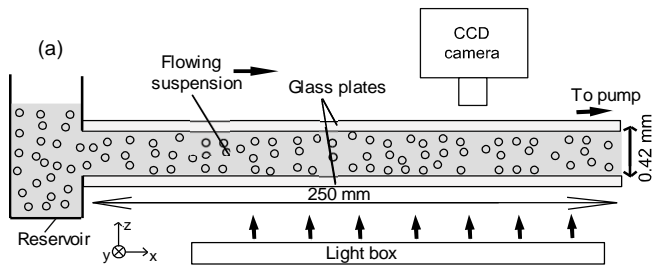

**Fig. 1**: Scheme of the experimental setup. Coordinates *x*,*y*,*z* correspond to the streamwise, spanwise and wall-normal directions respectively.

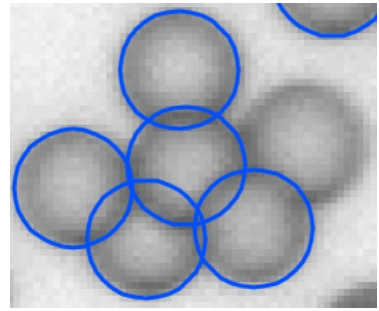

**Fig. 2**: Detection and resolution of groups of visually overlapping particles. The particle to the right of the image is shown without its detection circle as a reference to the reader.

The mean suspension velocity $U_s$ imposed by the pump is varied between 0.05 and 0.44 mm/s with an accuracy of 0.005 mm/s. Thus, the maximum particle Reynolds number was $Re_p = \frac{\rho a^2 \dot{\gamma}}{\mu} = 4.1 \cdot 10^{-2}$ and the minimum particle Péclet number was $Pe_p = \frac{6\pi \mu a^3 \dot{\gamma}}{kT} = 1.11 \cdot 10^5$, where $\dot{\gamma} = U_s/d$ is the characteristic shear-rate, $\rho$ is the density of the particles, $\mu$ is the fluid viscosity, $k$ is the Boltzmann's constant, $d$ is the channel thickness, and $T$ is the room temperature. The suspension was imaged using a CCD Camera located 2 cm above the channel with the optical axes normal to the parallel plates. The depth of field was such that it was possible to detect particles across the whole channel thickness. The zone under study had dimensions 1.05 by 0.8 mm, and was located at 180 mm from the inlet, where the steady-state suspension velocity profile is assumed as achieved. Indeed, it is known that the latter is achieved at a distance from the inlet (i.e. the entrance length) that scales as $d^3/a^2$ [18], which corresponds to 185 mm in the present work, assuming a scaling factor of 1. In the steady state, no net particle motion in the wall-normal direction (*z*) is expected. Images were acquired at 30 frames per second during 30 seconds, and typically, using a Hough transform algorithm, 500 particles were detected on each image. Taking into account the dimensions of the zone under study and the channel thickness, this was in good agreement with the target volume fraction $\phi = 0.05$ from the suspension preparation. Using a spatial resolution in which $a \approx 12$ pixels, we resolved groups of visually overlapping particles with rather good accuracy (Fig. 2). Trajectories were constructed using a minimal total square displacement rule and particle velocities were obtained using a second order scheme. Measurement statistics were improved by sampling velocities at consecutive time windows, separated by a few characteristic correlation times $t_c^*$ (to be defined later) during each experiment. As a consequence, the statistics shown in the figures typically imply $10^4$ velocity measurements.

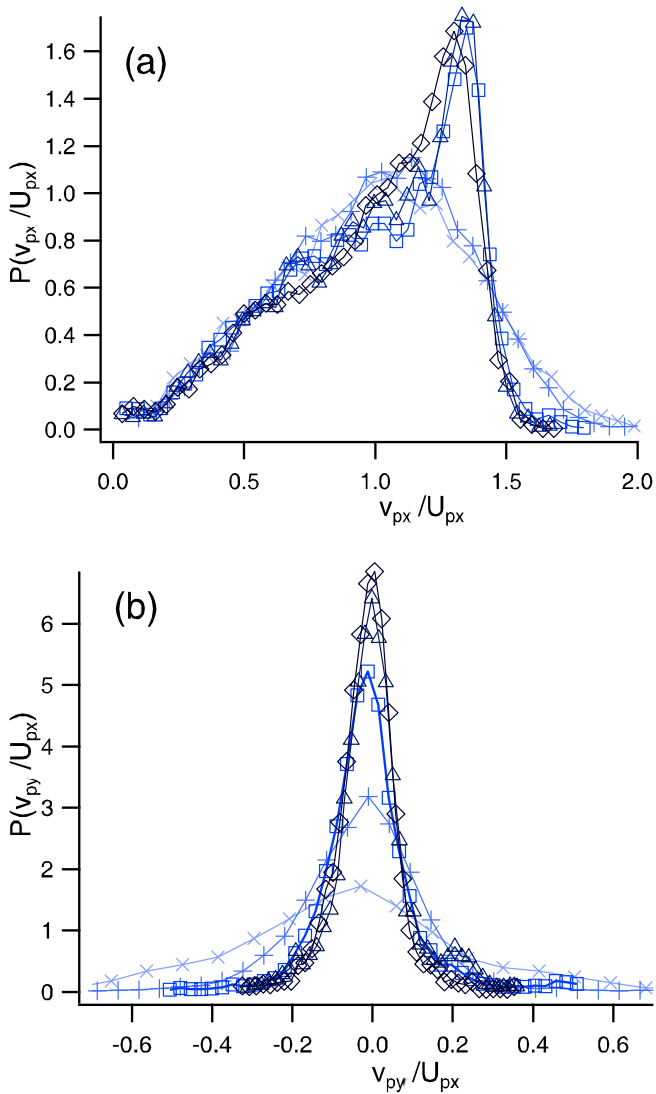

travelling at near-maximum velocity, as reported in the literature for similar experimental conditions [17-18]. However, a variation of the pdf with $U_s$ is surprising since the entrance length and the steady state suspension velocity profile are frequently assumed to be independent of $U_s$ [18]. It should also be noted that for $U_s < 0.174$ mm/s, the measurements yield some velocity values larger than the maximum one of the corresponding Poiseuille flow. This situation lacks physical sense, since the measured particle velocities should be smaller (if flattening occurs) or equal than the latter, but never greater. This discrepancy might be due to a certain degree of inaccuracy in the tracking procedure. Despite the mentioned discrepancy in Fig. 3(a), as one can see in Fig. 4(a) (displaying the mean and the maximum of the streamwise particle velocities), the maximum velocities satisfy the condition mentioned above within the experimental error, for all $U_s$ values.

Similarly, $U_{px}$ is in rather good agreement with $U_s$ (slightly greater as $U_s$ increases) and, although particle migration is evident in Fig. 3(a), it would correspond, as reported for a Brownian suspension at $\phi = 0.05$ [23], to a relatively weak change in the particles distribution in the thickness of the channel. The pdf of the spanwise ($y$) velocities remains Gaussian-like and narrows as $U_s$ increases. This suggests that the spanwise velocity fluctuations decrease with $U_s$.

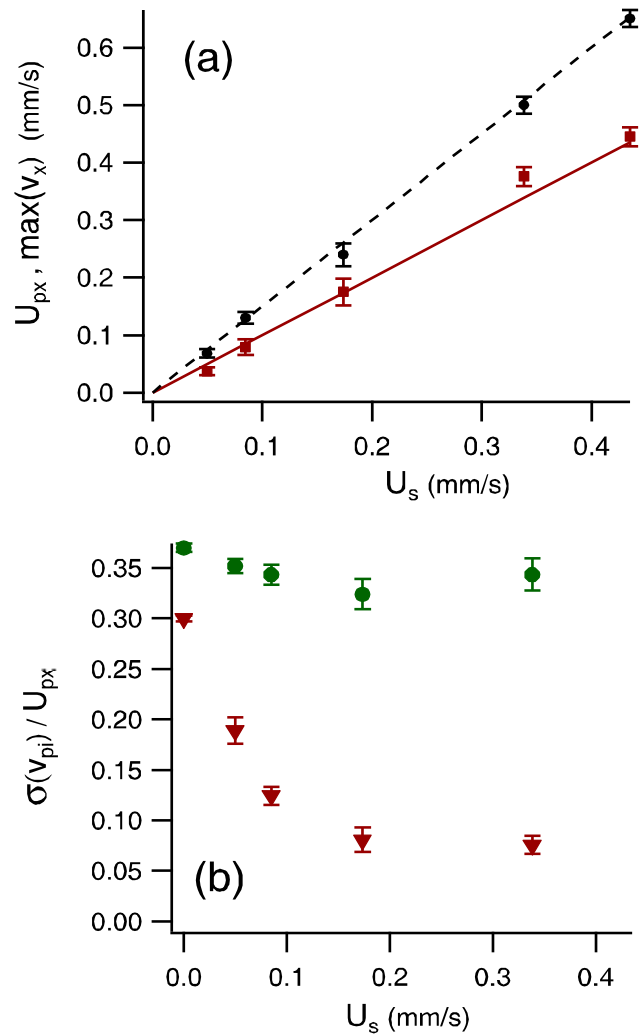

**Fig. 3**: Probability density functions (pdf) of the streamwise (*a*) and spanwise (*b*) particle velocities normalized by the mean streamwise velocity $U_{px}$. (×): $U_s$=0.05, (+): $U_s$ =0.085, (□): $U_s$ = 0.174, (∆): $U_s$=0.338, (◊): $U_s$ =0.435 mm/s. Streamwise (*x*), as $U_s$ reaches 0.174 mm/s, the shape of the pdf varies from symmetric to markedly asymmetric strongly skewed towards high velocities, with the most probable velocity close to the maximum velocity of a Poiseuille flow ($v_{px}/U_{px} \approx 1.5$). In the spanwise direction *y*, the pdf is always symmetric Gaussian-like, and narrows as $U_s$ increases. The solid lines are shown as guides to the eye.

## Results

Fig. 3 displays probability density functions (pdf) of the streamwise (Fig. 3(a)) and spanwise (Fig. 3(b)) particle velocities, normalized by the mean streamwise velocity $U_{px}$. Fig. 3(a) clearly exhibits a dependence of the pdf of the streamwise velocities with $U_s$. As $U_s$ increases, the pdf changes significantly, from symmetric, to asymmetric strongly skewed to high velocities, with a marked peak of the most probable velocity very close to the maximum velocity of a Poiseuille flow (= 1.5 $U_s$).

The latter behavior is a sign of particle migration towards the mid-plane of the channel, with a significant fraction of the particles

**Fig. 4**(a): Mean (■) and maximum (●) of the streamwise particle velocities as a function of $U_s$. Solid and dashed lines: predictions from Poiseuille flow between parallel plates. As $U_s$ increases $U_{px}$ varies from slightly smaller to greater than the Poiseuille prediction, while the maximum velocities are always smaller than the latter. (b): Streamwise (●) and spanwise (▼) velocity fluctuations as a function of $U_s$.

We confirmed this by calculating the velocity fluctuations $\sigma(v_{pi})/U_{px}$ from the velocities pdf, displayed on Fig. 4(b) (here $U_{px}$ is the mean of the streamwise velocity of the particles and $\sigma(v_{pi})$ is the standard deviation in the streamwise or spanwise direction). The magnitude of the spanwise velocity fluctuations decreases as $U_s$ increases in agreement with a migration of the particles towards the mid-plane of the channel. According to the suspension balance model and measurement of the second normal stress difference, streamwise and spanwise velocity fluctuations should be of the same order[19], so streamwise velocity fluctuations should also decrease with $U_s$. However, this is not observed in the present work, and streamwise velocity fluctuations remain rather constant with $U_s$, despite the sharp change in the pdf.

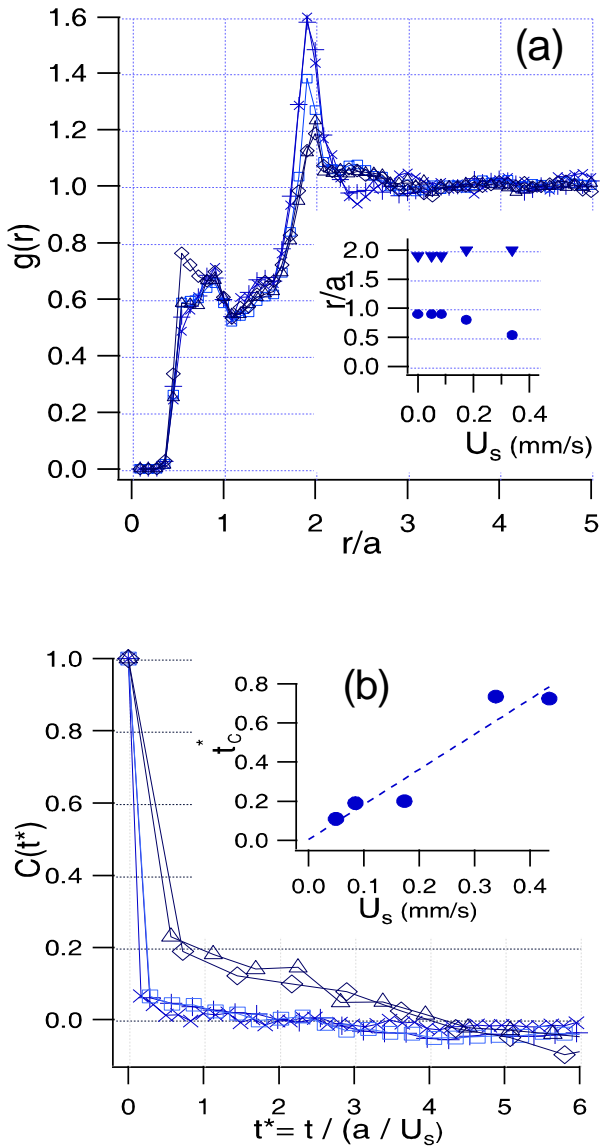

**Fig. 5(a):** Radial pair correlation function $g(r)$ where $r$ is the distance between particle centers measured in the *xy* plane of Fig 1. (×): $U_s$ =0.05, (+): $U_s$ =0.085, (□): $U_s$ = 0.174, (Δ): $U_s$ =0.338, (◊): $U_s$ =0.435 mm/s. Inset: Position of the first (●) and second (▼) peak as a function of $U_s$. Fig. 5 (b): Streamwise time autocorrelation functions $C(t^*)$ as a function of dimensionless time $t^*=t/(a/U_{px})$. The correlation increases as $U_s$ increases. By integrating $C(t^*)$ we obtained the dimensionless correlation time $t_c^*$ [25]. Spanwise, we observed oscillations from the first lag because our time sampling was too slow to resolve time correlation. The solid lines are shown as guides to the eye.

Suspension microstructure can be characterized by determining the pair correlation function $g(r)$. Fig. 5(a) shows the radial pair correlation function $g(r)$ where $r$ is the distance between particle centers measured in the *xy* plane of Fig 1. Two peaks are clearly visible for all values of $U_s$, their corresponding values of $r/a$ are plotted in the inset. A peak near $r/a= 2$ is characteristic of inhibited particle interpenetration due to solid boundaries (typically tightly packed spheres) and, in our configuration, it could be associated to particles organised in a clustering layer over the mid-plane. This peak adjusts to $r/a= 2$ as $U_s$ increases, indicating closer contact. The other peak is in the range $0.5 < r/a < 0.9$, which means that, in the camera view, particles are partially overlapped, with their centers apart. If particles were in contact, this peak would imply a preferential relative positioning angle between them of 20° with respect to the line of sight. Therefore, this peak may be related to the asymmetry of the pair correlation function due to the contact between particles. This asymmetry has been measured in viscosimetric flows for volume fraction as low as $\phi$=0.05 [24], and its variation with $U_s$ (inset Fig. 5(a)) indicates a gradual modification of the suspension microstructure.

Figure 5(b) shows the streamwise time autocorrelation function $C(t^*)$ (as defined in [25]) as a function of dimensionless time $t^*=t/(a/U_{px})$. For all values of $U_s$, the function becomes negative before tending to zero; some authors associated this behavior to that of a liquid structure [26]. For $U_s$ greater than 174 mm/s, $C(t^*)$ increases showing that particles move less independently from each other. The streamwise dimensionless correlation time $t_c^*= t_c/(a/U_{px})$ can be obtained as described in [25]. In the inset of Fig. 5(b), it is shown that $t_c^*$ increases with $U_s$. In the spanwise direction (*y*), $C(t^*)$ showed an oscillating behavior (statistical fluctuations) from the smallest possible lag, meaning that our time sampling was too poor to resolve $t_c^*$ in that direction.

The streamwise dimensionless diffusivity of the particles is finally calculated as
$$D_{xx}^* = t_c^* \left(\frac{\sigma(v_x)}{U_{px}}\right)^2 \equiv t_c^* \left(\sigma^*(v_x)\right)^2$$
[14], and is shown in Fig. 6 as a function of $U_s$. $D_{xx}^*$ increases with $U_s$ mainly due to an increase of $t_c^*$ (the velocity fluctuations remain rather constant). Despite this increase, we remark that the values of $D_{xx}^*$ are bounded by 0.1 for the range of $U_s$ studied.

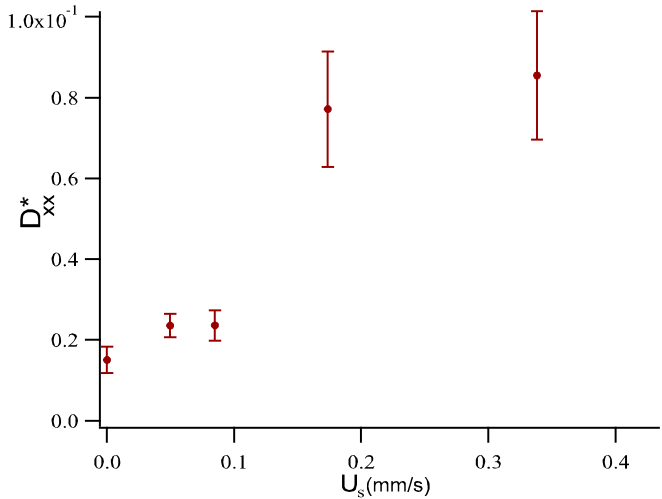

**Fig. 6**: Dimensionless particle diffusivities $D^*_{xx} = t_c^* \; (\sigma(v_{px})/U_{px})^2$ as a function of $U_s$. $D^*_{xx}$ increases with $U_s$ mainly due to the increase of $t_c^*$ (the velocity fluctuations remain rather constant), but remains smaller than 0.1 for the range of $U_s$ studied.

## Discussion

The results presented here are complementary to those of Roht et al.[7], in the sense that, the same phenomenon, under the same experimental conditions, in their work is studied at the scale of the spatial variation of the solute along the channel, and in ours is analysed at the scale of the suspended particles. We intended to investigate the mechanisms through which particles may affect the dispersion of the solute by comparing the results of both studies.

Roht et al.[7] observed a transition between a range of $U_s$ for which solute dispersivity was not affected by the presence of the particles, ($Pe_s < 168$, corresponding to $U_s < 0.19$ mm/s), and another one for which it clearly decreased as $\phi$ increased ($Pe_s > 283$, $U_s > 0.32$ mm/s). The decrease in the dispersivity of the solute then only occurs when $U_s$ is greater than a critical value $U_c \approx 0.19$ mm/s.

In our experiments, we observed that when $U_s$ exceeded $U_c$, a marked change in the pdf of the streamwise velocity of the particles took place (cf. Fig. 3(a)). This suggests that the decrease observed by *Roht et al.*[7] is related to a modification of the flow structure, in particular, to the onset of particle migration towards the mid-plane and consequent flattening of the velocity profile. The sign of migration is suggested by the appearance of a peak of most probable velocity very near the maximum velocity value in Fig. 3(a), that indicates a high fraction of particles travelling near the mid-plane, and also by data shown in Fig. 5(a), where the peak to the right of the figure adjusts to *r/a=2*, which suggests a closer contact between particles.

Regarding the mixing effect of the diffusive motion of the particles on the solute, we recall that we measured dimensionless particle diffusivities $D^*_{xx}$ of $O(10^{-1})$. Considering that $\sigma^*(v_y) << \sigma^*(v_x)$ (Fig. 4(b)), and that $t_c^*$ is much smaller in the spanwise direction $y$, then we may plausibly assume $D^*_{yy} << D^*_{xx}$, and it is also reasonable to expect that $D^*_{zz} << D^*_{xx}$. This is supported by previous results in other geometries, for instance, Breedveld et al.[27] reported spanwise (*y*) and wall-normal (*z*) diffusivity coefficients of the same order. In this context, particle diffusivities in all directions result of $O(10^{-1})$ or smaller.

In comparison, Roht et al.[7] measured dimensionless solute dispersion values $D^*_s$ of $O(10^3)$, that decreased almost 10% in the presence of the particles for $U_s > U_c$, being the decrease of $O(10^2)$. Under the strong assumption that particle random diffusive motion, characterized by a given particle dimensionless diffusivity (i.e. $D^*_{xx}$), imparts the solute a dimensionless diffusivity of the same order of magnitude, then the contribution of this motion is three orders of magnitude smaller than the reported decrease ($O(10^{-1}) << O(10^2)$). Independently of the mechanisms engendering the decrease, the mixing effect generated by particle motion can be then considered negligible.

Finally, for $U_s < U_c$, we don't find significant evidence of particle migration, but the spanwise velocity fluctuations are much larger than for greater values of $U_s$. According to the streamline distortion as a source of dispersion hypothesis, this situation would lead to enhancement, but this was not observed by Roht et al.[7]. However, it is possible that, if enhancement existed, the experimental device was not accurate enough to detect it.

## Conclusions

We experimentally studied particle the velocity statistics and the microstructure of a suspension flowing in a narrow parallel-plates channel. As the mean suspension velocity $U_s$ exceeded $U_c$, we observed a marked change the pdf of the streamwise velocity of the particles, with strong evidence of particle migration towards the mid-plane of the channel. This result sheds light on the decrease of the solute dispersivity measured by Roht et al.[7] for $U_s > U_c$ : particle migration leads to a flattening of the suspension velocity profile, in turn, this flattening involves a reduction of the velocity gradients in the wall-normal direction, making the solute molecular diffusion across the flow less effective.

We stress that the present analysis (in particular the estimation of particle diffusivities) does not capture the spatial correlation of the particle velocities, which could be important to characterize the flow structure. Unfortunately, the present technique is not suitable for obtaining a 3D description of particle motion (that might be however experimentally difficult to access in a highly confined flow configuration such as the one used in the present work).

Besides, we estimated the contribution of the random diffusive motion of the suspended particles on the dispersivity of the solute by calculating the diffusivity they impart to the latter. We found that this contribution is negligible compared with the decrease of the dispersivity experimentally measured by Roht et al.[7]. In short, solute dispersion is much more affected by how particles modify the spatial organisation of the flow, than by the mixing effect due to their random diffusive motion.

Nevertheless, in a different type of flow configuration, for example a microfluidic device, the influence of particle random diffusive motion on solute dispersivity might be non–negligible, and even important, the inclusion of particles potentially being an efficient method for enhanced mixing.

## Acknowledgements

We thank J. Brady, J.P. Hulin, L. Oger and F. Rouyer for helpful discussions and comments, A. Cánepa for help with data processing, and programs PIP CONICET 0246, PICT 20100802 and LIA PMF-FMF for support.